\documentclass[reprint,prl,nobibnotes,noshowpacs,superscriptaddress,floatfix,letterpaper]{revtex4-1}
\usepackage{amsmath,amssymb,amsbsy,amsfonts,amsthm}
\usepackage{bm,bbm}
\usepackage{graphicx}
\usepackage{color}
\usepackage{physics}
\usepackage[dvipsnames]{xcolor}
\usepackage[colorlinks=true,citecolor=Cerulean,linkcolor=RubineRed,urlcolor=Cerulean]{hyperref}
\usepackage{silence}
\usepackage{sidecap}
\sidecaptionvpos{figure}{t}
\usepackage{floatrow}

\begin{document}

\title{Critical fluctuations and slowing down of chaos}

\author{Moupriya~Das}
\affiliation{Department of Chemistry,\
  University of Massachusetts Boston,\
  Boston, MA 02125
}
\author{Jason~R.~Green}
\email[]{jason.green@umb.edu}
\affiliation{Department of Chemistry,\
  University of Massachusetts Boston,\
  Boston, MA 02125
}
\affiliation{Department of Physics,\
  University of Massachusetts Boston,\
  Boston, MA 02125
}
\affiliation{Center for Quantum and Nonequilibrium Systems,\
  University of Massachusetts Boston,\
  Boston, MA 02125
}

\begin{abstract}

Fluids cooled to the liquid-vapor critical point develop system-spanning
fluctuations in density that transform their visual appearance.  Despite the
rich phenomenology of this critical point, there is not currently an
explanation of the underlying mechanical instability. How do structural
correlations in molecular positions overcome the destabilizing force of
deterministic chaos in the molecular dynamics? Here, we couple techniques from
nonlinear dynamics and statistical physics to analyze the emergence of this
singular state. Our numerical simulations reveal that the ordering mechanisms
of critical dynamics are directly measurable through the hierarchy of
spatiotemporal Lyapunov modes. A subset of unstable modes softens near the
critical point, with a marked suppression in their characteristic exponents
reflecting a weakened sensitivity to initial conditions. Finite-time
fluctuations in these exponents, however, exhibit diverging dynamical
timescales and power law signatures of critical dynamics. Collectively, these
results are symptomatic of a critical slowing down of chaos that sits at the
root of our statistical understanding of the singular thermodynamic responses
at the liquid-vapor critical point.

\end{abstract}

\maketitle

Fluctuations are sovereign in critical
phenomena~\cite{Stanley88,*Goldenfeld92}.  Fluids at the liquid-vapor critical
point are not immune. These critical points~\cite{Smoluchowski08} are unique
instabilities that punctuate the space of thermodynamic states. First
established experimentally by Andrews~\cite{Andrews69}, the liquid-gas critical
point was given a molecular explanation shortly thereafter by van der
Waals~\cite{Sengers02}. Van der Waals' picture, now a paradigm in liquid state
theory~\cite{Hansen86,*Barrat03}, is that the repulsive forces largely
determine the structural arrangements of molecules in non-critical liquids, not
the attractive forces. Near the liquid-vapor critical point, however, their
roles reverse and the paradigm shifts~\cite{Widom67}; dynamical fluctuations
reach macroscopic magnitude and overrule molecular size, shape, and
interactions in dictating bulk behavior. These fluctuations are generated by
the deeply nonlinear dynamics of classical, critical fluids. Yet, the
relationship between microscopic dynamical instability and the thermodynamic
singularity has never been entirely clear~\cite{Berry00}.

While the field of critical phenomena continues to absorb increasingly diverse
systems~\cite{NielsenBM00}, the basic phenomenology is firmly
established~\cite{Stanley88,*Goldenfeld92}. Its taxonomy is built on scaling
and universality, the similar behavior of dissimilar systems. Despite the early
discovery of their critical points, fluids were somewhat resistant to
classification~\cite{Sengers02}. Simulations~\cite{Simeoni10} and
theory~\cite{PolandSD16} were, and continue to be, integral in providing
mechanistic insights, the location of the critical point, and estimates of
static critical exponents~\cite{JohnsonZG93,Caillol98,PotoffP98,WatanabeIH12}.
Through simulations, the classical atomistic dynamics of fluids are also known
to be chaotic~\cite{BosettiP14}, which is measurable through the machinery of
nonlinear dynamics. Nonlinear dynamics, which includes the notion of
deterministic chaos in its repertoire, has given insight into the physical
mechanisms of the jamming transition in granular materials~\cite{Banigan13},
self-organizing systems~\cite{GreenCGS13}, evaporating collections of
nuclei~\cite{ColonnaB99,BonaseraLR95}, and the phase changes of atomic
clusters~\cite{AmitranoB92,*WalesB94,*CalvoL98,*Calvo10}. Recently, the
dynamics of model spatially-extended systems have been assigned to known
dynamic universality classes~\cite{PazoLP16,PazoLP16b}. But, the dynamic
scaling of chaos in fluids near thermodynamic critical points has not yet been
explored. The recent steps to further coalesce statistical physics and
nonlinear dynamics reinvigorate the question of how fluids, specifically their
atomistic dynamics, fit within the phenomenological architecture of critical
phenomena.

\begin{figure}[bt!]
\centering
\includegraphics[width=1.0\columnwidth,angle=0,clip]{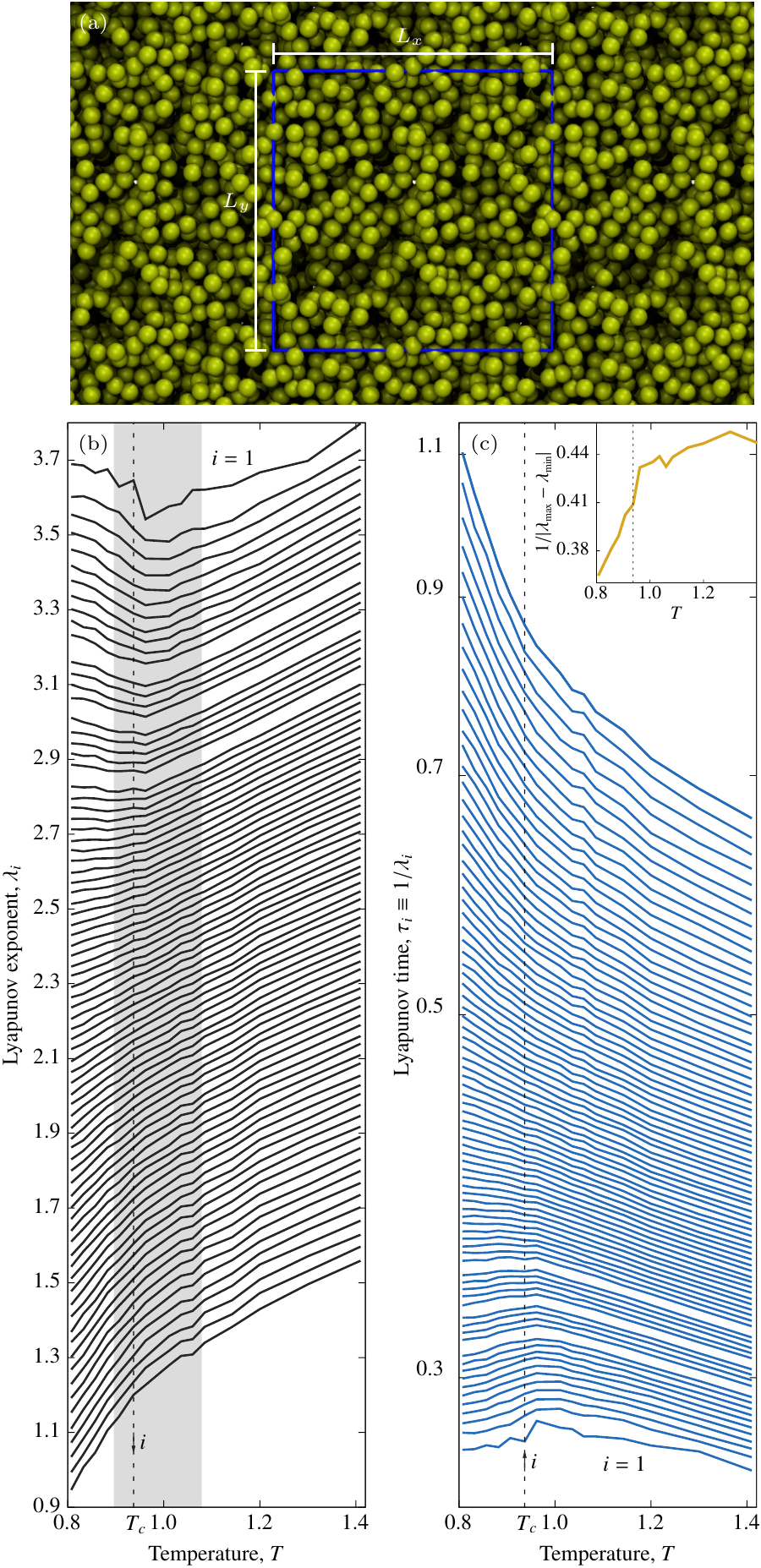}

\caption[]{\label{fig:fig1}Slowing down the divergence of trajectories at the
liquid-vapor critical point. (a) Snapshot of the critical fluid and the
periodic boundaries of length $L=L_x=L_y=L_z$. (b) Spectra of Lyapunov
exponents $\lambda_i$ and (c) Lyapunov times $\tau_i$ (log-linear scale) as a
function of the mean kinetic temperature, $T$. Every 30 $\lambda_i$ are shown
for $N=1000$ particles occupying a cubic simulation volume with a density
$\rho=\rho_c=0.317$. Unstable spatiotemporal modes that are more disordered
($1/3N\leq i/3N\leq 0.18$) have positive Lyapunov exponents (time) with a
minimum (maximum) in the supercritical region.  Vertical dashed lines mark the
critical temperature, $T_c$. Inset illustrates compression of spectrum through
the critical point for data shown.}

\vspace{-0.1in}

\end{figure}

At critical points, correlations span molecular-scale interactions to the
entire system. These correlations imply structural organization that is
intrinsically opposed by the chaotic dynamics. The mechanisms balancing this
internal tension between order and instability, however, are subtle. As a
result, theoretical explanations for the mechanical origins of critical
phenomena are uncommon. Continuous transitions in crystals are a notable
exception where structural changes arise through the instability of a lattice
vibration~\cite{Berry00}. There, the mode responsible for the phase transition
is a collective excitation whose frequency decreases anomalously during an
approach to the transition point. For example, in SrTiO$_3$ the frequency of a
soft phonon mode decreases substantially and, ultimately, freezes at the
transition temperature when approached from below~\cite{Scott74}. Because of
the absence of long-range order in fluids -- and an inability to make a small
vibration approximation, as in solids, or a molecular randomness hypothesis, as
in gases~\cite{LandauL58} -- the dynamical instability in critical fluids has
been grounded in purely statistical terms.

To analyze the instability of the liquid-vapor critical point, we apply
nonlinear dynamical techniques to molecular simulations of a homogeneous,
single-component, non-associated fluid. The system we simulate, shown in
Fig.~\ref{fig:fig1}(a), consists of $N$ particles interacting pairwise through
van der Waals forces, repelling (attracting) at distances of a (few) molecular
diameter(s) according to the Lennard-Jones potential~\cite{Jones24}. It has a
critical point, a triple point, and three phases (solid, liquid, and gas) that
are independently identifiable through a variety of molecular simulation
techniques~\cite{AllenT87,FrenkelS01} and is generally considered to belong to
the Ising universality class~\cite{Caillol98,WatanabeIH12}. We use the
hierarchy of $6N$ spatiotemporal modes, Lyapunov vectors, as order parameters.
Each mode has an associated exponent, $\lambda_{i}$ indexed $i=1,\ldots,
6N$~\cite{PikovskyP16} and in descending order, that measures the contribution
of each mode to the global dynamics. Each $\lambda_{i}$ is the average of the
$i$-th finite-time Lyapunov exponent $\lambda_i(t)$ over $10^4$ trajectories.
Larger exponents indicate more unstable, collective modes~\cite{TailleurK07}.
In our simulations at fixed number of particles $N$, energy $E$, and volume
$V$, we calculate the full Lyapunov spectrum at the critical density $\rho_c$
and over a range of temperatures including the critical temperature. We choose
the energy density to fix the mean kinetic temperature, $T$ (Methods).

\textit{The critical point as a limit of dynamical order in unstable Lyapunov
modes.--} When approaching the liquid-vapor critical point from above, $T>T_c$,
spatial regions form that will become vapor and liquid after the system is
cooled to $T<T_c$ through the critical point $T_c$. The statistical
correlations in these regions are also apparent in the Lyapunov modes, a set of
dynamic vectors representing collective changes to molecular positions and
momenta, Fig.~\ref{fig:fig1}~\cite{Gaspard98,PikovskyP16}. For example, the
magnitude of the largest Lyapunov exponent, $\lambda_1$, is dominated by the
fastest dynamical events in the system~\cite{PoschF2005}. In simple fluids,
the fastest events are pairwise interactions sampling the repulsive part of the
potential~\cite{DasCG17}. As shown in Fig.~\ref{fig:fig1}(b), there is a
minimum in the largest Lyapunov exponent near the known~\cite{ErringtonD03}
critical point, ($T_c,\rho_c$), giving direct evidence that critical conditions
inhibit the effect of repulsive forces on the dynamics. Compared to the
coexistence, $T<T_c$, or supercritical regimes, $T>T_c$, the minimum in this
exponent and the maximum in the Lyapunov time, $1/\lambda_1$ in
Fig.~\ref{fig:fig1}(c), also means the critical dynamics are predictable over
longer timescales because initially similar configurations do not diverge as
quickly. That is, the statistical correlations typically invoked to explain
this critical phenomenon are a reflection of the slowing down of chaos and
suppressed instability in critical dynamics.

\begin{SCfigure*}[][t]
\includegraphics[width=0.6\textwidth,angle=0,clip]{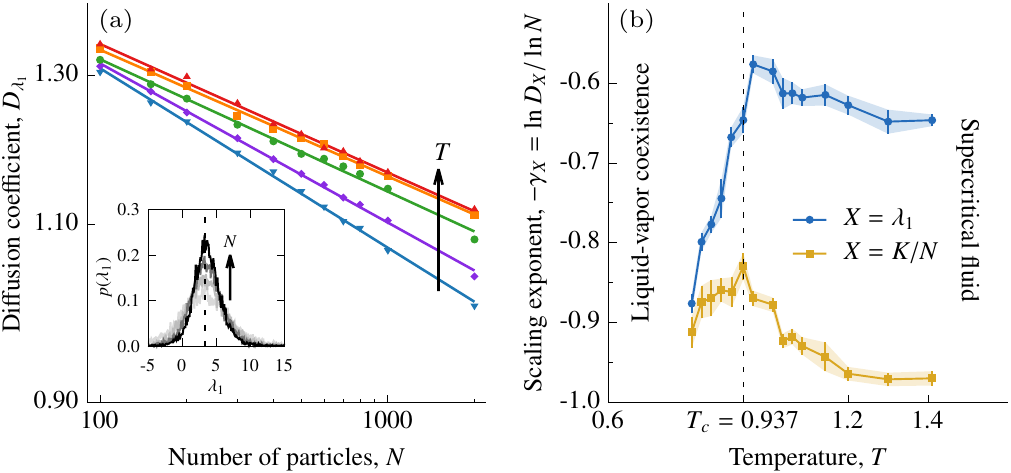}


\caption{\label{fig:fig3}Finite-size scaling of fluctuations in dynamical
observables. (a) Fluctuations in the kinetic energy and Lyapunov exponents
decay with increasing number of particles (log-log): the diffusion coefficients
follow a power law $D_{\scriptsize{X}}(N,T)\sim
N^{-\gamma_{\scriptsize{X}}(T)}$ for $11$ system sizes from $N=100$ to $2000$.
The corresponding probability distributions also concentrate (Inset: first
Lyapunov exponent at $\rho=\rho_{c}=0.317$ and $T=T_c=0.937$) (b) The wandering
exponents, $\gamma_{\scriptsize{X}}(T)$, of the first finite-time Lyapunov
exponent and the kinetic energy per particle, $K/N$, peak at the critical
temperature, $T_c$, at the critical density $\rho=\rho_{c}$. Error bars
indicate standard error ($1\,\sigma$) for parameter estimates from linear fits
in (a).}

\vspace{-0.2in}

\end{SCfigure*}

The minimum in the first Lyapunov exponent signals the breakdown of the van der
Waals picture, from which statistical mechanical treatments of liquids are
typically built~\cite{Widom67}. These perturbative treatments assume the
structure of a dense, monatomic liquid resembles that of a hard sphere fluid
and, to a first approximation, the attractive interactions have little effect
on the liquid structure~\cite{Barrat03}. However, this picture does not hold
near the critical point, and repulsive intermolecular forces instead play a
subordinate role compared to critical fluctuations as we see through the
suppression of the first Lyapunov exponent.  This finding differs from the
jamming transition in granular materials, though, which seems to be a
transition from a chaotic to a non-chaotic state~\cite{Banigan13}. The data in
Fig.~\ref{fig:fig1} suggests the continuous liquid-vapor phase transition
remains chaotic throughout and into the coexistence region for finite-size
systems.

To more fully resolve dynamical instability across the critical point, we
measure the full spectrum of Lyapunov exponents. There are $3N$ positive
exponents over the range of temperatures, suggesting the dynamics are chaotic
(Methods). The shape of the spectrum depends sensitively on kinetic temperature
and density~\cite{DellagoP96,YangR05,DasCG17}.  Fig.~\ref{fig:fig1}(b) shows
the spectrum at the critical density for temperatures spanning the critical
temperature, $T_c$. The exponents of all unstable modes decrease when
approaching the critical point from above. But, only the most unstable modes
have exponents with minima at the critical point.  Critical correlations appear
to have the largest impact on the modes with scaled index up to $i/3N \approx
0.18$. Exponents beyond this point increase monotonically with temperature. The
spectrum is also compressed at the critical point, Fig.~\ref{fig:fig1}(c)
inset, meaning there is a weaker preference for trajectories to diverge in the
direction of any given mode. 

Overall, these spectra show that critical dynamics are less sensitive to the
detailed features of intermolecular forces but also initial conditions. Because
the Lyapunov modes are mechanical objects, with their own equations of motion,
their associated spectra are direct evidence that these spatiotemporal modes
resolve the mechanical instability generating the liquid-vapor critical point.
Through these observables, critical conditions appear to constrain the dynamics
so that different phase space directions have a homogeneous degree of
instability relative to the supercritical $T>T_c$ or liquid-vapor coexistence
$T<T_c$ regimes.



\begin{SCfigure*}
\includegraphics[width=0.6\textwidth,angle=0,clip]{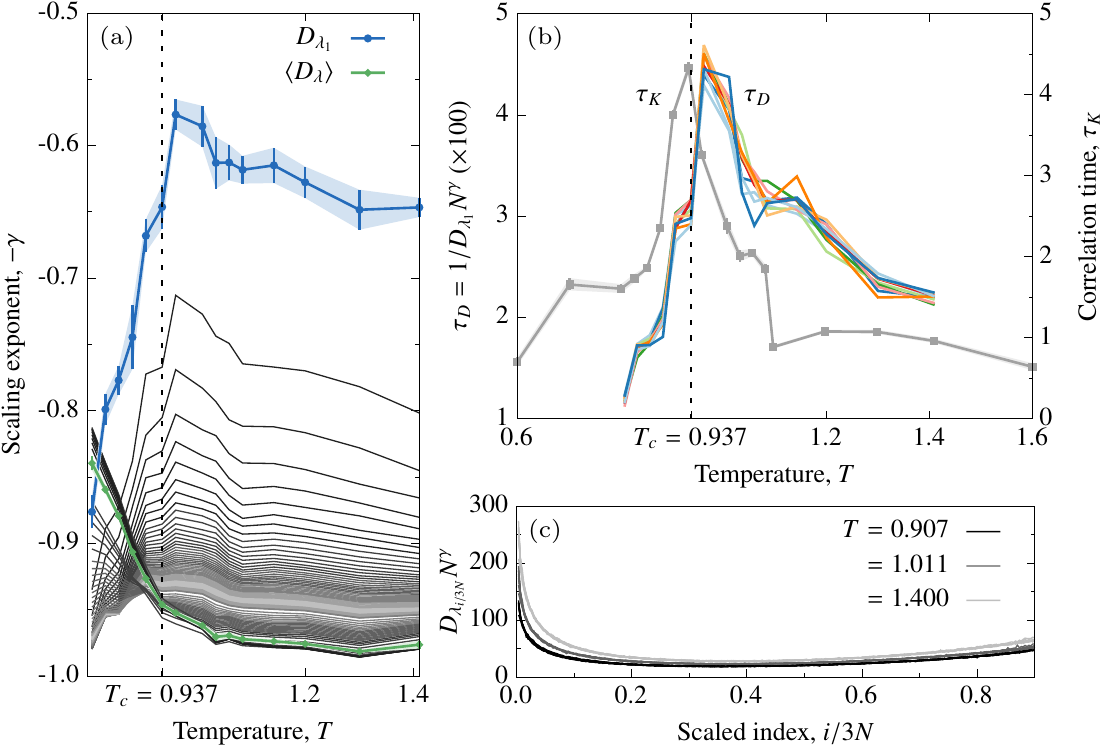}

\caption[]{\label{fig:fig4}Critical slowing down of dynamical time scales of
Lyapunov exponent fluctuations. Self-averaging and statistical dependence of
spatial domains for the first exponent and the bulk near the critical point.
(a) The (negative) wandering exponent as a function of temperature $T$ at
$\rho_{c}$ for the diffusion coefficient of the first finite-time Lyapunov
exponent. Also shown are the $-\gamma$ for average diffusion coefficients where
averaging includes an increasing number of Lyapunov exponents in the spectrum.
(b) Peak in the dynamical timescale for Lyapunov exponent fluctuations, $\tau_D
\equiv 1/D_{\lambda_1}N^\gamma$, and peak in the correlation time of kinetic
energy per particle, $\tau_K$. (c) Data collapse for the scaled diffusion
coefficient of finite-time Lyapunov exponent spectra as a function of the
scaled spectral indices: the functional form is the scaling function
$\tilde{D}(T,\rho=\rho_c)$. All data are at the critical density for the
three-dimensional Lennard-Jones fluid.}


\vspace{-0.1in}

\end{SCfigure*}

\textit{Scaling relations and diverging timescales in fluctuations.--} Cooling
the fluid towards the critical point causes the correlation length to diverge.
In our simulations, though, the correlation length saturates at the size of our
simulation cell, $L$. Fleeting clusters of all sizes up to $L$, which will
eventually become liquid, begin to appear.  These clusters' ephemeral existence
clearly suppresses chaos in the critical dynamics on long timescales. However,
their structure is only the spatial component of the mechanical system. The
Lyapunov modes, though, are local directions in position and momentum. As the
system structure evolves, the dynamics continues to temporarily sample phase
space domains where trajectories diverge more quickly and more slowly than the
average, domains that contribute significantly to the Lyapunov modes and the
finite-time estimates of the Lyapunov exponents.  Distributions of
$\lambda_{1}(t)$ are shown in Fig.~\ref{fig:fig3}(a).

Power laws are a hallmark of critical phenomena~\cite{Callen85} that quantify
the divergence of a system's response to small changes in external fields such
as temperature. Non-trivial~\cite{AharonyH96} power law behavior is apparent in
the decay of temporal Lyapunov exponent fluctuations with system size. At
thermodynamic equilibrium, the precise scaling of relative fluctuations, termed
self-averaging~\cite{Milchev1986}, in an observable $X$ decay according to
$R_X=\left<\Delta X^2\right>/\left<X\right>^2 \sim
D_{\scriptsize{X}}/\left<X\right>^2 \sim N^{-\gamma}$ with $\Delta X =
X-\left<X\right>$, the wandering exponent $\gamma$, and a generalized diffusion
coefficient $D_{\scriptsize{X}}$ (SI Sec.~III). Loosely speaking, a system is
self-averaging with respect to a given property $X$ if the value of the
thermodynamic observable corresponds to the average over independent subsystems
or, in this case, time windows. More precisely, the relative variance $R_{X}$
of the property $X$ vanishes in the thermodynamic limit: $R_X\to 0$ when
$N\to\infty$. The wandering exponent $\gamma$ can have values between $0$ and
$1$ -- a value of one meaning the observable is strongly self-averaging. Weakly
self-averaging observables have $\gamma <1$ and non-self-averaging observables
have $\gamma = 0$. Because of the statistical independence of spatial domains
in the system, equilibrium observables are often strongly self-averaging with
$\gamma=1$. However, these domains become statistically dependent near a
critical point because of the divergence of the correlation
length~\cite{Stanley88,Goldenfeld92}. Statistical signatures of dynamical
observables, like the Lyapunov exponents, are still being
elucidated~\cite{PazoLP16,PazoLP16b}. Deep in the liquid state, for example,
the fluctuations for all but the first Lyapunov exponent (the ``bulk'') are
strongly self-averaging. The first exponent fluctuations, however, self-average
weakly with a rate of decay $\gamma < 1$ that depends on the length scale of
the interparticle interactions and captures the van der Waals picture of
dominant attractive forces~\cite{DasG17}. How does this picture change at the
critical point?

To quantify the scaling behavior of the finite-time Lyapunov exponent
fluctuations with system size, we simulate $11$ systems ranging from $N=100$ to
$2000$ over the same range of temperatures at fixed density $\rho_{c}$.
Numerical calculation of the entire Lyapunov spectrum comes at significant
computational cost, cost that increases significantly when scaling with system
size~\cite{CostaG13}. Fluctuations in the first finite-time Lyapunov exponent,
$\lambda_1(t)$, as measured by the diffusion coefficient
$D_{\lambda_{1}}(N,T,\rho)$, decay with system size as $N^{-\gamma}$ at
temperatures $T=0.8-1.4$ spanning the critical point, Fig.~\ref{fig:fig3}(a).
Temperature, through the structural correlations it brings about, strongly
affects the wandering exponent, $\gamma$, which varies between $0.9$ to $0.6$.
A higher-order statistical analysis indicates this weak self-averaging is, at
least in part, due to the non-Gaussian features of the distributions (SI
Sec.~IV). A weakened, albeit significant, decay in this three-dimensional fluid
is distinct from the self-averaging found so far in one and two-dimensional
dynamical systems where long-range correlations cause fluctuations to
diverge~\cite{DasG17,PazoLP16}.

Most prominent in the temperature dependence of the wandering exponent of the
first exponent is the peak at $T=0.962$, Fig.~\ref{fig:fig3}(b). It shows the
critical dynamics self-average most weakly in the direction of the first
Lyapunov vector, and again indicating the sensitivity of this unstable mode to
the dominant length scales in the system. As a reference, the fluctuations in
kinetic energy per particle also decay with system size; the wandering exponent
peaks at $T=0.937$, confirming the location of $T_c$ and further quantifying
the statistical dependence of spatial domains, Fig.~\ref{fig:fig3}(b). This
temperature agrees with that found from grand canonical Monte Carlo
simulations~\cite{ErringtonD03}. However, the peak in $\gamma$ for the first
Lyapunov exponent is just above the critical temperature (and at the same
temperature as the minimum in the long-time average $\lambda_{1}$). A likely
explanation is the rounding of the peak caused by the finite-size of the system
that is typical in simulations of observables at the critical point. We also
note that extrema are known in the long-time Lyapunov exponents of simple
fluids near phase boundaries. For example, the sum of positive exponents
exhibits a maximum near, but not at, the fluid-to-solid phase-transition
density~\cite{PoschF2005}. Other information theoretic quantities are also
known to peak on the disorder side of an order-disorder transition, as was
shown in a kinetic Ising model~\cite{Barnett13}. What is clear from these
maxima we see in the wandering exponent is that the long-range correlations at
the critical point stabilize the most unstable mode on long time scales and
sustain large fluctuations in this direction on short timescales. Together,
these wandering exponents of fluids under critical conditions add to the
mounting evidence that the extending length scale of intermolecular
correlations can weaken the self-averaging of global fluctuations in dynamical
observables~\cite{DasG17,PazoLP16}. 

In many cases, the self-averaging of the first finite-time Lyapunov exponent is
distinct from that of the bulk, the set of $3N-1$ exponents that exclude
the first. In liquids, all the bulk exponents are strongly
self-averaging~\cite{DasG17}. However, the critical point, breaks this scaling
symmetry -- a significant fraction of the bulk exponents self-average weakly as
shown in Fig.~\ref{fig:fig4}. Even for spatially-extended dynamical systems
with long-range correlations and diverging fluctuations, the scaling of the
bulk of the spectrum is homogeneous~\cite{PazoLP16}, so this statistical
feature is so far unique to the liquid-vapor critical point. It is also
apparent in the self-averaging of the entire Lyapunov spectrum through the
average diffusion coefficient $\langle D(\lambda) \rangle = \sum_{i=1}^{6N}
D(\lambda_{i})/6N$, which contrasts that of the largest exponent,
Fig.~\ref{fig:fig4}(a). The corresponding wandering exponent has an inflection
point around $T=0.962$, again, at the critical density. Increasing the fraction
of exponents included in the average shows that a portion of the more unstable
modes have a muted $\gamma$-peak that vanishes with increasing index.

While fluctuations appear to decay with system size in all unstable phase space
directions on our accessible time and length scales, the rate of decay is far
from homogeneous. The $\gamma$-spectrum quantitatively resolves the rates at
which this unique thermodynamic equilibrium state emerges from the atomistic
dynamics, Fig.~\ref{fig:fig4}(a). The good data collapse in
Fig.~\ref{fig:fig4}(c) reveals clear scaling functions for the diffusion
coefficient spectrum, $\tilde{D}(T,\rho=\rho_c)$; only three representative
temperatures are shown.  This scaling function is a system-size independent
measure of the finite-time fluctuations in the Lyapunov spectrum, fluctuations
caused by the local heterogeneities in phase space sampled by our simulated
trajectories. The basic form of this scaling function is similar to that seen
for simple liquids~\cite{DasG17}. Our calculation of $D_{\lambda_1}$ leads
directly to a dynamical timescale for $\lambda_1(t)$ fluctuations, $\tau_D =
1/D_{\lambda_1}N^\gamma$. This timescale peaks just above the critical
temperature~\cite{ErringtonD03}, which suggests that in a critical state,
fluids experience amplified fluctuations along the most unstable mode, making
the dynamics easier to predict over longer times. We take this signature to be
another symptom of critical slowing down of chaos.  Although these modes are
highly active on short timescales, the fluctuations destructively interfere on
longer timescales (showing a net suppression, Fig.~\ref{fig:fig1}, that is
strongest at the same temperature).  The peak in the correlation time of the
kinetic energy per particle is evidence of traditional critical slowing down,
Fig.~\ref{fig:fig4}(b).

In summary, by treating the nonlinear dynamics directly, we have resolved the
collective spatiotemporal modes responsible for the thermodynamic instability
and the breakdown of the van der Waals picture at the liquid-vapor critical
point. Their observable properties exhibit universal features and scaling
distinctive of critical dynamics. Unlike continuous crystal-crystal phase
transitions, there is not one unique unstable mode. Instead, the whole spectrum
softens with a subset with extrema near the critical point; both the long-time
Lyapunov exponents and their fluctuations on short times reflect their high
sensitivity to long-range correlations of molecular positions. These modes do
not appear to completely freeze, at least not in finite-size systems, but do
exhibit large fluctuations with a peak in dynamical timescales indicating the
critical slowing down of chaos, the stabilization of unstable modes, and a
longer memory of initial conditions. In short, the relative \textit{mechanical
stability} of molecular motion underlies the bulk behavior of fluids at this
\textit{thermodynamic instability}.

\textit{Methods.--} Our system is the three-dimensional, periodic Lennard-Jones
fluid. The Hamiltonian of this $N$-particles is
$H(r_{lm},p_{k})=\sum_{k}^{3N}p_{k}^{2}/2m+\sum_{l<m}^{N}V(r_{lm})$, where
$V(r_{lm})$ denotes the interparticle interactions between the particles $l$
and $m$ at $r_{lm}$ distance apart. The pairwise interactions between particles
have short-range repulsions and comparatively longer range attractions. All
quantities are in reduced units (SI, Sec.I).

At fixed number of particles $N$, volume $V$, and energy $E$, we simulate
deterministic trajectories of this equilibrium fluid and the dynamics of
Lyapunov vectors in tangent space~\cite{CostaG13}.  The $i$-th Lyapunov vector
components are first variations in position and momentum $(\delta q_{ij},
\delta p_{ij})^{T}$ with $i,j = 1,\ldots, 6N$ and evolve according to their own
linearized Hamiltonian equation of motion. We numerically solve this equation
of motion with the linearized form of the velocity Verlet algorithm, used to
evolve trajectories, and orthonormalization at every time step~\cite{CostaG13}.
The initial basis sets are random and orthonormal.  During a transient, that we
discard, the first vector orients itself parallel to the maximally changing
tangent space direction. Regular orthonormalization restricts the collapse of
the remaining vectors onto the most expanding tangent space dimension. The
algorithm requires the second derivatives (Hessian) of the interaction
potential at every time steps. We have used forward differences of the
analytical gradients with a displacement of $10^{-4}$.

Within the linearized limit, the expansion or the compression factor along
the phase-space direction of the $i$-th Lyapunov vector over time $t$ is
$e^{\Gamma_{i}(t)}$. The corresponding finite-time Lyapunov exponent is
$\lambda_{i}=\Gamma_{i}(t)/t$.  The complete finite-time Lyapunov spectrum,
$\{\lambda_{i}\}$ is calculated from the set of Gram-Schmidt vectors with
standard methods~\cite{Fujisaka83,BenettinGGS80}.  We evaluate the full
Lyapunov spectrum at each time step using the metric $\left|\delta
x_{i}\right|^{2} =\sum_{j}^{6N} [\delta q_{ij}(t)^{2}+\delta p_{ij}(t)^{2}]$.
The $i$-th finite-time exponent over a time interval $\Delta t = t-t_0$ has the
form, 
\begin{eqnarray}
  \lambda_{i}(\Delta t)= \left|t-t_0\right|^{-1} \ln \frac{\left|\delta x_{i}(t) \right|}{\left|\delta x_{i}(t_{0})\right|}. \nonumber
\end{eqnarray}

Finite-time Lyapunov exponents are fluctuating variables.  We estimate the
magnitude of their fluctuations over a time intervals, $t$, with the diffusion
coefficients $\{D(\lambda_{i})\}$~\cite{PazoLP16,PazoLP16b,DasG17} and the
variance, $\chi_{i}^{2}(t)$, of $\{\Gamma_{i}(t)\}$
\begin{eqnarray}
tD(\lambda_{i}) = \chi_{i}^{2}(t) = \left<(\Gamma_{i}(t)- \left<\lambda_{i} \right>t )^{2} \right>. \nonumber
\end{eqnarray}
Averages $\left<\cdot\right>$ are over an ensemble of $10^{4}$ trajectories,
each having time span $t=0.1$ in reduced time units (SI Sec.I).  The average
$\left<\lambda_{i}\right>$ is the average of the $i$-th Lyapunov exponent over
an entire trajectory. To probe the self-averaging property of finite-time
Lyapunov exponent fluctuations, we ran trajectories over a range of system
sizes. We scaled the number of molecules $N$ and the volume $V$ to ensure the
thermodynamic limit of the microcanonical ensemble: $N,V\to\infty$ keeping the
number density $\rho=N/V$ and energy density $e=E/V$ constant.  According to
the equipartition theorem, the kinetic temperature of the system is given by
$T=2\left<E_\textrm{kin}\right>/3Nk_\textsc{\tiny B}$.  We analyze the scaling
of fluctuations in dynamical variables with system-size $N$ for a range of $T$
(SI Sec. III and IV) with $\rho$ fixed at $\rho_{c}=0.317$.

\begin{acknowledgments}

Acknowledgment is made to the donors of The American Chemical Society Petroleum
Research Fund (ACS PRF $\#$ 55195-DNI6) for support of this research. We
acknowledge the use of the supercomputing facilities managed by the Research
Computing Group at the University of Massachusetts Boston. We thank Bala
Sundaram for helpful comments on the manuscript.

\end{acknowledgments}


%

\end{document}